\documentclass[11pt]{article}

\date{}  

\usepackage{url}
\usepackage{cite}
\usepackage{plain}

\usepackage{wrapfig}
\usepackage{graphics}
\usepackage{picins,graphicx}
\usepackage[english]{babel}
\usepackage{amssymb}
\usepackage{verbatim}
\usepackage[mathscr]{eucal}
\usepackage{amstext}
\usepackage{makeidx}
\usepackage{amsthm}  
\usepackage{amsmath}

\usepackage{hyperref}
\hypersetup{colorlinks,
citecolor=red,
filecolor=blue,
linkcolor=blue,
urlcolor=blue}

\makeindex

\newtheorem{theorem}{Theorem} 

\newcommand{\mail}[1]{\href{unina:#1}{\texttt{#1}}}

\usepackage{pdfsync}

\author{Monica De Angelis*, Gaetano Fiore   \thanks{Univ. of Naples  "Federico II", Faculty of Engineering, Dip. Mat. Appl. "R.Caccioppoli", \newline
 Via Claudio n.21, 80125, Naples, Italy. \mail{modeange@unina.it}}}
 
\title{Diffusion effects in  a superconductive model}

\begin{document}
\maketitle

\begin{abstract}
A superconductive model characterized by a third order parabolic operator $ {\cal L}_\varepsilon  $  is analysed.  When the viscous terms, represented by higher - order derivatives, tend  to zero, a hyperbolic operator $ {\cal L}_0 $ appears. Furthermore,  if   ${\cal P}_\varepsilon$  is  the Dirichlet  initial 
boundary - value problem for $ {\cal L}_\varepsilon$,
when ${\cal L}
_\varepsilon  $ turns into  ${\cal
L}_0 , $
   ${\cal P}_\varepsilon$ turns into a problem  ${\cal P}_0$  with the {\em
same} initial - boundary conditions as  ${\cal P}_\varepsilon$. The solution of the nonlinear problem related to the remainder term $ r $  is achieved,  as long as the higher-order derivatives of the solution of ${\cal P}_0$  are bounded. Moreover, some classes of  explicit solutions related to $ {\cal P}_0 $  are determined, proving the existence of at least one  motion whose derivatives are bounded. The estimate shows that the diffusion effects are bounded even when time tends to infinity. 

\end{abstract}

\vspace{3mm}

{\bf AMS} {35K35,35E35}

\vspace{3mm}


\section{Introduction }

We deal  with the following  third order parabolic  equation:

\begin{equation}   \label{11}
(\partial_{xx} \, - \,\lambda\,  \partial _x\,)\,\,(\varepsilon
u _{t}+ u) - \partial_t(u_{t}+\alpha\,u)\,=F(x,t,
  u)
\end{equation} 
 which  characterizes  a lot of phenomena  (see, for instance, \cite{mwascom2011} and references therein). In particular, when $ F=\sin u -\gamma  \,\,( \gamma= const), $ the semilinear equation (\ref{11})  describes the evolution of the phase $ \,u\, $ inside  an exponentially shaped Josephson junction (ESJJ)\cite{bcs96,j005}.

 The (ESJJ)  provides several advantages in comparison with  rectangular junctions and  the  related bibliography is large and extensive.\cite{mwascom2011,bcs00,bcs96,cmc02,cwa08,bss,j05,j005,ssb04}. Besides, interesting  applications of   Josephson junctions are  SQUIDs which can be  applied in a lot of fields \cite{100}. Indeed, they are very versatile and they are  used in  medicine for  the Magnetoencephalography (MEG) and for Magnetic resonance imaging (MRI) or in cosmology, beeing SQUIDs   extremely sensitive detectors of magnetic and
electric signals from any origin. 
Moreover, in geophysics they are used  as magnetometers to discover mineral deposits, or as  X-ray detectors
to screen  radioactive materials.\cite{c,c2}. Furthermore, they are	     suitable in the development of smart materials such as superconductor cables,  used to	    improve power grids.\cite{bn,ml}

 Equation  (\ref{11})  is  equivalent  to the equation:

 \begin{equation}   \label{12}
\varepsilon  u_{_{xxt}} -  u_{tt} +\, u_{xx}\,-\,(\alpha\,+\, \frac{\varepsilon\,\lambda^2}{4•})\,u_t\,\,=\,\, \frac{\varepsilon \lambda^2}{4•} \,u+\,e^{\,-\,\frac{\lambda}{2•}\,x}\,F(x,t,\,e^{\frac{\lambda}{2•}\,x}
 u)
\end{equation}

 \noindent and from this,  letting

\[  a= \alpha\,+\, \frac{\varepsilon\,\lambda^2}{4•}- \frac{1}{\varepsilon•}\quad  \,\,\mbox{ and} \quad  \,\,\, b= \, \frac{\,\lambda^2\,\varepsilon\,\,}{4•} -\frac{\,a}{\varepsilon•},
\]

\noindent   it is possible to obtain the following system :

\begin{equation}     \label{13}
  \left \{
   \begin{array}{lll}
    \displaystyle{\frac{\partial \,u }{\partial \,t }} =\,  \varepsilon \,\frac{\partial^2 \,u }{\partial \,x^2 }
     \,-\, v\,\, -a \,u  - \varepsilon\,e^{-\,\frac{\lambda}{2•}\,x} \,\,F(x,t, e^{\frac{\lambda}{2•}\,x}
  u) \,  \\
\\
\displaystyle{\frac{\partial \,v }{\partial \,t } }\, = \, b\, u\,
- \frac{1}{\varepsilon•}\, v\, - \varepsilon \,\,u_t\,F_t\,(x,t,e^{\frac{\lambda}{2•}\,x}  u) 
\\
   \end{array}
  \right.
 \end{equation}

  \noindent which  is related to biological dynamical systems such as, for instance,   the Fitzhugh Nagumo model  (FHN) which reproduces the neurocomputational  features of the  neuron \cite{i,m1}.

\noindent So, as it has already been  underlined in \cite{mda2010,acscott,acscott02,dr08,r} , by means of  equation (\ref{11}) it is possible to treat both  superconductive problems and   neuroscientific ones.

In this paper,  the Dirichlet initial boundary value problem  for equation (\ref{11}) is considered and,   according to the theory of the phenomena related to (ESJJ), as  source term, it is  assumed that $ F=\sin u -\gamma  \,\,( \gamma= const) $.

Equation (\ref{11}) is a semilinear hyperbolic equation, perturbed by viscous terms which are described by higher - order derivatives with small diffusion coefficients $ \varepsilon $.   When  $ \varepsilon  \,\equiv 0, $   the parabolic equation $(\ref{11})$ turns into a  hyperbolic equation:

\begin{equation}  \label{14}
(\partial_{xx} \, - \,\lambda\,  \partial _x\,)\,u_0 - \partial_t(\partial_t+\alpha\,)u_0\,= F(x,t,u_0),\
\end{equation}

    \noindent and the influence of the dissipative terms $\, \varepsilon \,( u_{xxt}\,- \lambda u_{xt})$ on the wave behaviour of $\,u_0\,$ can be  estimated when   the difference

\begin{equation}                \label {15}
 r(x,t,\varepsilon)\,= u(x,t,\varepsilon)  - u_0(x,t)  
\end{equation}

 \noindent  is  evaluated.

A similar problem has already been studied in \cite{md}, when the coefficient $ \lambda $ in (\ref{11}) is equal to zero and for finite  intervals of time. Moreover, in \cite{dmr,dr1}, for  $\lambda =0 \,\,\,\mbox{and }  a=0, $ an asymptotic approximation is established by means of the  two characteristic times: slow time $ \tau = \varepsilon \,t $  and fast time $ \theta =\, t/\varepsilon $.  

 In this paper, in order to evaluate the diffusion effects, a rigorous estimate of the reimander term $ r, $  defined in (\ref{15}), is done. Indeed,  by means of a Fourier series with properties of rapid convergence, the solution in both linear and non linear case of the problem related to $ r $, is determined. Furthermore, some classes of solutions of the hyperbolic equation have been determined, proving that  there exists almost a solution whose  derivatives are  bounded. Finally, as soon as  the hyperbolic equation admits  solutions with limited derivatives, an estimate for the remainder term   is   achieved by proving that the  diffusion effects    are both  bounded when $ \varepsilon $ goes to zero and  when the time $ t $ tends to infinity.

\section{Statement of the problem}

If $T, \,\,l$  are two  positive constants and 

\vspace{4mm}

$\ \ \ \ \ \ \ \ \ \ \ \  \ \ \ \ \ \ \   \Omega_T =\{(x,t) :  0<x<l 
, \  \ 0 < t \leq T \}$,

\noindent  let us  consider the  initial boundary   value problem ${\,\cal
P}_\varepsilon \,$:

  \begin{equation}          \label{21}
  \left \{
   \begin{array}{ll}
    & (\partial_{xx} \, - \,\lambda\,  \partial _x\,)\,\,(\varepsilon
u _{t}+ u) - \partial_t(u_{t}+\alpha\,u)\,=F(x,t,u),\ \  \
       (x,t)\in \Omega_T,\vspace{2mm}\\
   & u(x,0)=h_0(x), \  \    u_t(x,0)=h_1(x), \  \ x\in [0,l],\vspace{2mm}  \\
    & u(0,t)=\varphi_0, \  \ u(l,t)=\varphi_1, \  \ 0<t \leq T.
   \end{array}
  \right.
 \end{equation}

\noindent where $\, F,\,\, \,h_{i},\,\, \varphi_{i}\, (i=0,1)$   are arbitrary specified functions and $  \varepsilon, \, a, \, $  and   $\lambda$ are positive constants.

If  $ \varepsilon  \,\equiv 0, $   the parabolic equation $(\ref{21})_1$ turns into the  hyperbolic equation (\ref{14})  and the new problem  ${\cal
P}_0 $ for $ u_0(x,t)$ has {\em the same initial boundary  conditions  of }$\,\,{\cal
P}_\varepsilon. \,$\vspace{3mm}

 \noindent To evaluate the influence of the dissipative terms, letting 
 
 \begin{equation}                   \label{22}
\,\,\bar{F}(x,t,r)= F(x,t,r+u_0) - F(x,t,u_0) - \, \varepsilon \, \partial_{xt}\,( \partial_x\,- \lambda) u_0,\,\,
\end{equation} 
 
  \noindent the  following  problem  ${\,\cal
P}\,$   related to the {\em  remainder} term $ \, r ,\,$  defined in (\ref {15}) has to be considered:

\begin{equation}          \label{23}
  \left \{
   \begin{array}{ll}
    & (\partial_{xx} \, - \,\lambda\,  \partial _x\,)\,\,(\varepsilon
\partial_t \,+1)\,r - \partial_t(\partial+\alpha\,)r\,=\bar F(x,t,r),\ \  \
       (x,t)\in \Omega_T,\vspace{2mm}\\
   & r(x,0)=0, \  \    r_t(x,0)=0, \  \ x\in [0,l],\vspace{2mm}  \\
    & r(0,t)=0, \  \ r(l,t)=0, \  \ 0<t \leq T.
   \end{array}
  \right.
 \end{equation}

As for the model of superconductivity, the problem (\ref{23}) is characterized  by the source force: 

\begin{equation}  \label{24}
\,\bar F \,= \, \sin  \, (r+u_0) \, -\sin u_0  \,- \, \varepsilon \, \partial_{xt}\,( \partial_x\,- \lambda) u_0.   
\end{equation}

\section{The Green function and its properties}

Let us introduce the linear third order operator: 

 \begin{equation}                                   \label{31}
 {\cal L}_ \varepsilon =(\partial_{xx}-\lambda \partial_x)(\varepsilon
\partial_t+1 ) - \partial_t(\partial_t+\alpha).
 \end{equation}

  \noindent  The Green function of ${\cal L}_ \varepsilon  $ has  already been determined  in \cite{mwascom2011} by means of Fourier method. Let

\begin{equation}                                             \label{32}
\gamma_n=\frac{n\pi}{l},\  \quad  b_n=(\gamma_n^2\,+ \lambda^2/4\,), \quad  g_n=\frac{1}{2} (\alpha\,+\,\varepsilon\,b_n\,),\  
\  \omega_n=\sqrt{g_n^2- \,b_n\,}
\end{equation}

\noindent and

\begin{equation}                                \label{33}
G_n(t)= \,\,\frac{1}{\omega_n}\,\,e^{-g_n\,t}\,\,
sinh(\omega_nt),
\end{equation}

 \noindent standard techniques allow to  obtain   the Green function  as:

\begin{equation}                                                 \label{34}
G(x,t,\xi)=\frac{2}{l}\,\, e^{\frac{\lambda\,}{2\,}\,x}\,\,\sum_{n=1}^{\infty}\,
G_n(t) \  \  sin\gamma_n\xi\  \ sin\gamma_nx.
\end{equation}

The following theorem, proved in \cite{mwascom2011},  ensures  that this series is characterized by properties of rapid convergence and that it is exponentially vanishing as $ t $ tends to zero.

\begin{theorem}   \label{t21} 
 \noindent Whatever the constants $\,\,\alpha,\,\, \varepsilon,\,\, \lambda\,\, $
 may be in $\,\Re^+\,$, denoting  by $ \, a_\lambda = \alpha \, + \varepsilon \, \lambda^2/4 \,\, $ and with


\begin{equation}                      \label{36}
p_\lambda= \frac{\pi^2}{\varepsilon\,\pi^2 \,+\,a_\lambda\,l^2},  \ \ \  q_\lambda=  \frac{\,a_\lambda\,+\,\varepsilon(\pi/l)^2}{2}, \ \ \ \delta\equiv min(p_\lambda, q_\lambda),
\end{equation}

\noindent the function $G(x,\xi,t)$ defined in
(\ref{34})  and all its time derivatives, are continuous functions in
$\Omega_T$ and it results:

\begin{equation}                                        \label{37}
|G(x,\xi,t)|\,\leq \,M  \,e^{-\delta t}, \ \ \, \ \ \ \
|\frac{\partial ^j G}{\partial t^j}|\, \leq \,N_j \,  e^{-\delta
t},
\\ \ \ \ \ \ \, \ \ j \in {\sf N}
\end{equation}

\noindent  where  $M, N_j $ are constants depending on  $\alpha, \lambda, \varepsilon$.

Moreover, one has:

\begin{equation}                            \label{38}
 |\partial_{x}^{(i)}\,\,(\varepsilon\, G_t\, +\, G )|\,\,\leq A_i \, \,
\, e^{-\delta t},\qquad (i=0,1,2) 
\end{equation}

\noindent
 where $A_i\,\,\,(i=0,1,2)$   are constants depending  on $a, \varepsilon ,\lambda $. $\hfill\square $\end{theorem}

The uniform convergence proved in theorem \ref{t21} allows  to prove that \cite{mwascom2011}:

\begin{equation}                      \label{39}
{\cal L}_\varepsilon G \, =(\partial_{xx}\, -\, \lambda \, \partial_x)(\varepsilon
G _{t}+ G) - \partial_t(G_{t}+\alpha\,G)=0.
\end{equation}

\section{ The remainder term}
  
 To make an  estimate of  the solution to the  problem  ${\cal P} $ related to the remainder term $ r, $  two cases should be distinguished according to  the source  term being  known or depending by the  unknown function.

Let us assume  $ \bar F(x,r,t) =f(x,t).$  In this circumstance the remainder term $ r $ can be expressed by  an explicit form. Indeed,  standard computations (see f.i. \cite{ddr98}) ensure that  the solution of the linear problem (\ref{23}) is given by

\begin{equation}                                          \label{41}
 r_f(x,t)=\,-\,
\int_0^ t d\tau\, \int_0^l\, G(x,\xi,t-\tau)\,
f(\xi,\tau)d\xi.
\end{equation}

\noindent Otherwise, if the source term $ \bar F $ is non linear, an integro differential formulation of the problem can be obtained. 

 \noindent Let us introduce the following 

{\textbf{Conditions A}:  Let us  assume that function $\,F(\,x,t,u\,)\, $  is defined and continuous on the set

\begin{equation}  \label{42}
  \,\,\, D_T \ \equiv  \{ \, (x,t,u)\,\,: \,(x,t)  \in   \Omega_T \,, \,\,-\infty \,<\,u\,<\infty \,  \}
 \end{equation}

\noindent and  that it is   Lipschitz  continuous in $ \, \,u\, \,$  i.e.  there exists a constant $\,K\,$ such that the inequality

\begin{equation} \label{43}
\, |F (x,t,u_1)\,-\,F (x,t,u_2)|\, \leq \,\,K \,\, \ \, | u_1-u_2\,|  
\end{equation}

 \noindent  holds for all $\, (\,u_1,\,u_2\,)$ and $ (\,x,\,t\,) \, \in \Omega_T  $.

Under Assumption A, by means of the fixed point theorem, it is possible to prove \cite{dmm,daf,df} that the non linear problem ${\,\cal
P} \,$ admits a solution given by

\begin{equation}                                          \label{44}
 r(x,t)=\,
\int_0^ t d\tau\, \int_0^l\, G(x,\xi,t-\tau)\,
\bar F(\xi,\tau,r(\xi,\tau))d\xi.
\end{equation}

\section{Explicit solutions of the hyperbolic equation }
 
Let us consider the semilinear  second order equation:  

\begin{equation}                \label{51}
  u_{0,xx} \,- \,u _{0,tt} \, - \, \ \alpha \,u_ {0,t} - \lambda \,u_{0,x} =   \sin u_0 \,  \,- \gamma  
\end{equation}

When $ \lambda \, = \, 0  $,  (\ref{51}) represents  the  perturbed sine - Gordon equation (PSGE) and  there is plenty of  literature   about its   classes of solution. (e.g.\cite{ddf,ds,gh} and bibliography therein.)

Now, let $\, \psi \,$ be  an arbitrary function and let  us  consider $ \,\,\Pi (\psi)\,\,$  as:

\begin{equation}     \label{52}
\Pi(\psi) = \, 2 \, \arctan\,\, e^\psi \,\,  
\end{equation}

\noindent such that

\begin{equation}                \label{53}
  \sin \,  \Pi (\psi) \,\, = \frac{1}{\cosh (\psi) }, \qquad  \cos \,  \Pi (\psi) \,\, = -   \tanh (\psi) .   
\end{equation}

\noindent By means of function (\ref{52}) it is possible to find a class of solutions of  equation  (\ref{51}).

 \noindent Indeed, it is possible  to  verify   that the following function:

\begin{equation}
u_0 \, = 2 \, \Pi [f(\xi)]\,  \qquad \mbox{with}\qquad \xi =\, \frac{x-t}{\alpha-\lambda}
\end{equation}

 \noindent is a solution of (\ref{51}) provided that one has:

\begin{equation}  \label{55}
(\lambda-\alpha)\, u_{0,t}\, = \, \sin u_0 \, -\, \gamma. \, \,\,\end{equation}

\noindent Moreover, since (\ref{53}) and   beeing $\,\, \dot \Pi = \dfrac{1}{\cosh f}, \,\,\,  $ it results:

\[ (\lambda-\alpha)\, u_{0,t}\, = 2\frac{f'}{\cosh f}\]
\[ \sin u_0 \,\,=  2 \sin \Pi \cos \Pi  \,\,= \,-\,\,2 \,\frac{\tanh f }{\cosh f}, 
\]

\noindent  so, from (\ref{55}), one deduces that function $ f $ must satisfy the following equation:

\begin{equation}  \label{56}
\frac{df}{\tanh f\, +\gamma/2\, \cosh f•}\,= \,-d\,\xi.
\end{equation} 

\noindent Since  physical experiences show that generally $ 0\leq \gamma \leq 1,$ we point our attention to those cases in  which it results: 

\begin{equation}
u_0\,=\, 4 \arctan (\, y\,+\,\sqrt{y^2 \, +\, 1}\,).
\end{equation}

\noindent So, let $ h $  be an arbitrary constant of integration, one obtains:

\begin{equation}
 y= \,h\, e^{-\, \xi\,}\,\,\,\,\quad\quad \qquad \qquad \mbox{when}\qquad \qquad \gamma = 0,
\end{equation}

\begin{equation}
 y= \,\frac{1-(\xi\,-h)}{1+ \xi -h}\qquad \qquad  \mbox{when}\qquad \qquad \gamma = 1.
\end{equation}

\noindent Moreover, assuming $ \gamma <1,   $ let \[ A = \, \pm \sqrt{1-\gamma^2}\,\, \qquad \mbox{and} \qquad  \beta = 1+A.\]

\noindent  For   $ \gamma^2 \neq \beta^2, $ it results 

\begin{equation}  \label{510}
 y= \,\frac{h\, \frac{\gamma}{\beta\,\,}\,\,e^{\xi A}\,\,-\frac{\beta}{\gamma\,\,}}{1-h \, e^{\,\,\xi \,\,A}}. 
 \end{equation}

\noindent \textbf{ Remark 1}: In  the case  $ \gamma \,= \,0,  $   if  denoted by 

\begin{equation} \label{511}
 \,z\,(x,t)=\, e^{\,-\,\,\frac{x-t}{\alpha-\lambda} }+\, \sqrt{ e^{\,-\,\, 2\frac{x-t}{\alpha-\lambda} } \,+1}, \,\, 
 \end{equation}

\noindent equation (\ref{51}) admits the solution

\begin{equation} \label{512}
u_0(x,t)\,= \,4 \arctan\,[ z(x,t)].
\end{equation}
 
\noindent and it is possible to prove that

\begin{equation}  \label{513}
  \left \{
     \begin{array}{lll}           
u_{0,xt}(x,t)\,= \displaystyle{ \frac{1}{ 2 (\alpha-\lambda)^2 }\,\,}
 \frac{y\, ( \,y^2\,- 1 \,)} {( \, y^2 +1)^2}\\
\\ u_{0,xxt}(x,t)\,= \displaystyle{\frac{1}{ 2 (\alpha-\lambda)^3}\,}\,\, \,\frac{\,\, y\,\,(\, y^6\,  \,-\,5\, y^5\, -\,5\, y^2\,+1 )}{(y^2\,\,+1)\,^4\,} 
\end{array}
  \right. 
\end{equation}

 \noindent which are bounded for all $ (x,t) \in \Omega_T. $ 

\section{Estimates for the reimander term}

Let us assume  $ \varepsilon =0,  $  and let $ u_0  $  be  a solution of the problem ${\cal
P}_0.  $

\begin{theorem} \label{remainder}
If there exist   two  positive costants   $ h,\,k  $ depending only on $ \alpha  $ and $ \lambda $ such that 

\begin{equation} \label{61}
|u_{0,xt}(x,t)\,| \leq \,h; \qquad |u_{0,xxt}(x,t)|\, \leq \,k, 
\end{equation}

\noindent denoting by

\begin{equation}
S(t)\, = \sup_{\Omega_T} \, | r(x,t,\varepsilon)|, 
\end{equation}

\noindent the following inequality  holds:

\begin{equation}   \label{63}
0\,\leq \,S(t)\,\leq\,\displaystyle{\frac{ M\,l \,\,(h+k)\,\varepsilon \,}{\delta} \, } \,\,\, e^{\,\,(\,1\,-\,e^{-\,\delta \,t}\,)}\
\end{equation}

\noindent beeing  $ M $   a positive constant depending on   $\alpha, \lambda, \varepsilon$ and $ \delta  $ is defined by  (\ref{36}).
\end{theorem}
Proof :  Let us consider function  $\bar F $ defined in (\ref{24}):

\begin{equation}  \label{}
\,\bar F \,= \, \sin  \, (r+u_0) \, -\sin u_0  \,- \, \varepsilon \, \partial_{xt}\,( \partial_x\,- \lambda) u_0.   
\end{equation}

\noindent That   source term satisfies conditions A   and so, by means of  hypotheses (\ref{61}), it results:

\begin{equation}
|r(x,t,\varepsilon)| \,\leq \int_0^ t d\tau\, \int_0^l\, G(x,\xi,t-\tau)\,
[\,\,|r(\xi,t,\varepsilon)\,| + \varepsilon (h+k)\,]\,\,d\xi.
\end{equation}

\noindent By properties of the  Green function $G,  $ and in particular since $(\ref{37})_1, $ one obtains:

\begin{equation} \label{66}
0\,\leq \,S(t)\,\leq\, M\,l \, \int_0^ t \,  e^{\,-\,\delta \,(t-\tau)\,}\,
[\,S(\tau)\,+ \varepsilon \,(h+k)\,]\,d\tau
\end{equation}

\noindent So, by means of Gronwall Lemma, inequality (\ref{63}) holds.  $\hfill\square$

 \noindent   Inequality (\ref{63}) shows that, under assumptions of theorem  \ref{remainder},  in  the evolution of  $ u(x,t) $ the correction   due to the $ \varepsilon  $ - terms  is   bounded   even when t tends to infinity, and it  vanishes as soon as $ \varepsilon $ tends to zero.

\vspace{3mm} \textbf{Remark 2} Formulas (\ref{513}) show that  the class of functions    satisfying  hypoteses  of Theorem \ref{remainder}  is not empty.

\vspace{5mm}   
{\bf Acknowledgement}: This  work has been performed under the auspices of Programma F.A.R.O. (Finanziamenti per l' Avvio di  Ricerche
Originali, III tornata) ``Controllo e stabilita' di processi diffusivi nell'ambiente'', Polo delle Scienze e Tecnologie, Universita' degli Studi di Napoli Federico II. (2012)

\end{document}